\documentstyle[twoside]{article}
\input amssym

\catcode`\@=11
\long\def\@makefntext#1{
\protect\noindent \hbox to 3.2pt {\hskip-.9pt  
$^{{\eightrm\@thefnmark}}$\hfil}#1\hfill}		

\def\@makefnmark{\hbox to 0pt{$^{\@thefnmark}$\hss}}	
	
\def\ps@myheadings{\let\@mkboth\@gobbletwo
\def\@oddhead{\hbox{}
\rightmark\hfil\eightrm\thepage}   
\def\@oddfoot{}\def\@evenhead{\eightrm\thepage\hfil
\leftmark\hbox{}}\def\@evenfoot{}
\def\sectionmark##1{}\def\subsectionmark##1{}}

\oddsidemargin=\evensidemargin
\newcounter{sectionc}\newcounter{subsectionc}\newcounter{subsubsectionc}
\renewcommand{\section}[1] {\vspace{12pt}\addtocounter{sectionc}{1} 
\setcounter{subsectionc}{0}\setcounter{subsubsectionc}{0}\noindent 
	{\tenbf\thesectionc. #1}\par\vspace{5pt}}
\renewcommand{\subsection}[1] {\vspace{12pt}\addtocounter{subsectionc}{1} 
	\setcounter{subsubsectionc}{0}\noindent 
	{\bf\thesectionc.\thesubsectionc. {\kern1pt \bfit #1}}\par\vspace{5pt}}
\renewcommand{\subsubsection}[1] {\vspace{12pt}\addtocounter{subsubsectionc}{1}
	\noindent{\tenrm\thesectionc.\thesubsectionc.\thesubsubsectionc.
	{\kern1pt \tenit #1}}\par\vspace{5pt}}
\newcommand{\nonumsection}[1] {\vspace{12pt}\noindent{\tenbf #1}
	\par\vspace{5pt}}

\newcounter{appendixc}
\newcounter{subappendixc}[appendixc]
\newcounter{subsubappendixc}[subappendixc]
\renewcommand{\thesubappendixc}{\Alph{appendixc}.\arabic{subappendixc}}
\renewcommand{\thesubsubappendixc}
	{\Alph{appendixc}.\arabic{subappendixc}.\arabic{subsubappendixc}}

\renewcommand{\appendix}[1] {\vspace{12pt}
        \refstepcounter{appendixc}
        \setcounter{figure}{0}
        \setcounter{table}{0}
        \setcounter{lemma}{0}
        \setcounter{theorem}{0}
        \setcounter{corollary}{0}
        \setcounter{definition}{0}
        \setcounter{equation}{0}
        \renewcommand{\thefigure}{\Alph{appendixc}.\arabic{figure}}
        \renewcommand{\thetable}{\Alph{appendixc}.\arabic{table}}
        \renewcommand{\theappendixc}{\Alph{appendixc}}
        \renewcommand{\thelemma}{\Alph{appendixc}.\arabic{lemma}}
        \renewcommand{\thetheorem}{\Alph{appendixc}.\arabic{theorem}}
        \renewcommand{\thedefinition}{\Alph{appendixc}.\arabic{definition}}
        \renewcommand{\thecorollary}{\Alph{appendixc}.\arabic{corollary}}
        \renewcommand{\theequation}{\Alph{appendixc}.\arabic{equation}}
        \noindent{\tenbf Appendix#1}\par\vspace{5pt}}
\newcommand{\subappendix}[1] {\vspace{12pt}
        \refstepcounter{subappendixc}
        \noindent{\bf Appendix \thesubappendixc. {\kern1pt \bfit #1}}
	\par\vspace{5pt}}
\newcommand{\subsubappendix}[1] {\vspace{12pt}
        \refstepcounter{subsubappendixc}
        \noindent{\rm Appendix \thesubsubappendixc. {\kern1pt \tenit #1}}
	\par\vspace{5pt}}

\newcommand{\textlineskip}{\baselineskip=13pt}
\newcommand{\smalllineskip}{\baselineskip=10pt}

\def\eightcirc{
\begin{picture}(0,0)
\put(4.4,1.8){\circle{6.5}}
\end{picture}}
\def\eightcopyright{\eightcirc\kern2.7pt\hbox{\eightrm c}} 

\newcommand{\copyrightheading}[1]
	{\vspace*{-2.5cm}\smalllineskip{\flushleft
	{\footnotesize Mathematical Models and Methods in Applied Sciences #1}\\
	{\footnotesize $\eightcopyright$\, World Scientific Publishing
	 Company}\\
	 }}

\def\abstracts#1#2#3{{
	\centering{\begin{minipage}{4.5in}\baselineskip=10pt\footnotesize
	\parindent=0pt #1\par 
	\parindent=15pt #2\par
	\parindent=15pt #3
	\end{minipage}}\par}} 

\def\keywords#1{{
	\centering{\begin{minipage}{4.5in}\baselineskip=10pt\footnotesize
	{\footnotesize\it Keywords}\/: #1
	 \end{minipage}}\par}}

\renewenvironment{thebibliography}[1]
	{\frenchspacing
	 \ninerm\baselineskip=11pt
	 \begin{list}{\arabic{enumi}.}
        {\usecounter{enumi}\setlength{\parsep}{0pt}     
	 \setlength{\leftmargin 12.7pt}{\rightmargin 0pt} 
         \setlength{\itemsep}{0pt} \settowidth
	{\labelwidth}{#1.}\sloppy}}{\end{list}}

\newcounter{itemlistc}
\newcounter{romanlistc}
\newcounter{alphlistc}
\newcounter{arabiclistc}

\newcommand{\fcaption}[1]{
        \refstepcounter{figure}
        \setbox\@tempboxa = \hbox{\footnotesize Fig.~\thefigure. #1}
        \ifdim \wd\@tempboxa > 5in
           {\begin{center}
        \parbox{5in}{\footnotesize\smalllineskip Fig.~\thefigure. #1}
            \end{center}}
        \else
             {\begin{center}
             {\footnotesize Fig.~\thefigure. #1}
              \end{center}}
        \fi}

\newcommand{\tcaption}[1]{
        \refstepcounter{table}
        \setbox\@tempboxa = \hbox{\footnotesize Table~\thetable. #1}
        \ifdim \wd\@tempboxa > 5in
           {\begin{center}
        \parbox{5in}{\footnotesize\smalllineskip Table~\thetable. #1}
            \end{center}}
v        \else
             {\begin{center}
             {\footnotesize Table~\thetable. #1}
              \end{center}}
        \fi}

\def\@citex[#1]#2{\if@filesw\immediate\write\@auxout
	{\string\citation{#2}}\fi
\def\@citea{}\@cite{\@for\@citeb:=#2\do
	{\@citea\def\@citea{,}\@ifundefined
	{b@\@citeb}{{\bf ?}\@warning
	{Citation `\@citeb' on page \thepage \space undefined}}
	{\csname b@\@citeb\endcsname}}}{#1}}

\newif\if@cghi
\def\cite{\@cghitrue\@ifnextchar [{\@tempswatrue
	\@citex}{\@tempswafalse\@citex[]}}
\def\citelow{\@cghifalse\@ifnextchar [{\@tempswatrue
	\@citex}{\@tempswafalse\@citex[]}}
\def\@cite#1#2{{$\null^{#1}$\if@tempswa\typeout
	{IJCGA warning: optional citation argument 
	ignored: `#2'} \fi}}

\def\pmb#1{\setbox0=\hbox{#1}
	\kern-.025em\copy0\kern-\wd0
	\kern.05em\copy0\kern-\wd0
	\kern-.025em\raise.0433em\box0}

\def\fnt#1#2{\footnotetext{\kern-.3em
	{$^{\mbox{\scriptsize #1}}$}{#2}}}

\font\tenrm=cmr10
\font\tenit=cmti10 
\font\tenbf=cmbx10
\font\bfit=cmbxti10 at 10pt
\font\ninerm=cmr9

\font\eightrm=cmr8

\def\qed{\hbox{${\vcenter{\vbox{			
   \hrule height 0.4pt\hbox{\vrule width 0.4pt height 6pt
   \kern5pt\vrule width 0.4pt}\hrule height 0.4pt}}}$}}

\def\theequation{\thesectionc.\arabic{equation}}	

\def\be{\begin{equation}}
\def\ee{\end{equation}}
\def\t{\hbox}
\def\pr{\prime}
\def\p{\varphi}
\def\pp{\varphi^\pr}
\def\ppp{\varphi^{\pr\pr}}
\def\ps{\varphi^{\pr*}}
\def\a{\alpha}
\def\b{\beta}

\def\l{\lambda}
\def\q{\quad}
\def\f{\frac}
\def\il{\int\limits}
\begin{document}

\normalsize\textlineskip
\thispagestyle{empty}
\setcounter{page}{1}
\copyrightheading{}			

\vspace*{0.88truein}

\centerline{\bf CONTINUOUS ANALOG OF THE GAUSS-NEWTON METHOD}

\vspace*{0.37truein}
\centerline{\footnotesize 
RUBEN G. AIRAPETYAN
\footnote{E-mail: airapet@math.ksu.edu},\quad
ALEXANDER G. RAMM
\footnote{E-mail: ramm@math.ksu.edu}, \quad
and ALEXANDRA B. SMIRNOVA
\footnote{E-mail: smirn@math.ksu.edu} 
}
\vspace*{0.015truein}
\centerline{\footnotesize\it Department of Mathematics, Kansas State University,} 
\baselineskip=10pt
\centerline{\footnotesize\it Manhattan, Kansas 66506-2602, U.S.A.}

\vspace*{0.225truein}

\vspace*{1.0truein}

\vspace*{0.21truein}
\abstracts{
A Continuous Analog of discrete Gauss-Newton Method (CAGNM)
for numerical solution of nonlinear problems is suggested. In order to avoid
the ill-posed inversion of the Fr\'echet derivative operator some regularization
function is introduced. For the CAGNM a convergence theorem is proved. The
proposed method is illustrated by a numerical example in which a nonlinear
inverse problem of gravimetry is considered.  Based on the results of the numerical
experiments practical recommendations for the choice of the regularization function
are given.
}
{}{}



\vspace*{1pt}\textlineskip	
\section{Introduction}	        
\vspace*{-0.5pt}
\noindent
\setcounter{equation}{0}
\renewcommand{\theequation}{1.\arabic{equation}}

Let $H_1$ and $H_2$ be real Hilbert spaces and $\p:H_1 \to H_2$
a nonlinear operator. Let us consider the equation:
\be\label{eq1} \p(x)=0.\ee
We assume that the following condition on $\p$ holds.

\medskip
\noindent
{\bf Condition A}:\ 
Problem (\ref{eq1}) has a solution $\hat{x}$,  not necessarily unique.

In the well-known Newton's method (\cite{or}) one constructs a sequence $\{x_n\}$ for
$n=0,1,\dots$ which converges to a solution (in general non unique) of
equation (\ref{eq1}). The first term $x_0$ is an initial
approximation point and the other terms are constructed by means of the
following iterative process:
\be\label{nm}
x_{k+1}=x_k-\pp(x_k)^{-1}\p(x_k),
\ee
where $\pp(x)$ is the Fr\'echet derivative of the operator $\p$. Recall that
$\pp(x)$ is a linear operator from $H_1$ to $H_2$. The usual
necessary condition for the realization of the Newton method is the bounded
invertibility of $\pp(x_k)$, that is, the existence of a bounded linear
operator $[\pp(x_k)]^{-1}$ for all $k$. Actually in order to provide the convergence
of Newton iterations one needs bounded invertibility of $\pp$ in a ball
$B(\hat{x},R):=\lbrace x:x \in H_1, ||x-\hat{x}||\leq R\rbrace$.
However this condition does not hold in many important applications. In order
to avoid this restriction several modifications of the Newton method have been
developed. In this paper we consider the Gauss-Newton procedure for
equation (\ref{eq1}) (see e.g. (\cite{or})):
\be\label{eq2}
x_{k+1}=x_k-[\ps(x_k)\pp(x_k)]^{-1}\ps(x_k)\p(x_k), \q x_0=x_0.
\ee
If the operator $\ps(x)\pp(x)$ is not boundedly
invertible one needs some regularization procedure. In order to construct
such a procedure one can introduce a sequence of positive numbers
$\a_k$, $\a_k\to 0$, and replace iterative method~(\ref{eq2}) by the following one
(\cite{or,bak}):
\be\label{eq3}
x_{k+1}=x_k-[\ps(x_k)\pp(x_k)+\a_k I]^{-1}[\ps(x_k)\p(x_k)+\a_k(x_k-x_0)], 
\ee
where $I$ is the identity operator.

The methods constructed above can be also considered as discrete analogs
of some continuous methods (called sometimes continuation
methods). In (\cite{g}) the following Cauchy problem
has been considered as a continuous analog of (\ref{nm}):
\be\label{cnm}
\dot{x}(t)=-\pp(x(t))\p(x(t)),\q x(0)=x_0,  \q \dot{x}(t):=
\frac { dx }{dt}.
\ee
A solution to problem (\ref{eq1}) can be obtained as limit of the
function $x(t)$ for $t\to\infty$.
If one solves this Cauchy problem by means of Euler's 
method with a stepsize $\tau=1$ one gets (\ref{nm}) 
with $x_k=x(k)$. Continuous
analogs of iterative methods have several advantages over
the discrete ones. Convergence theorems for continuous methods usually
can be obtained easier. If a convergence theorem is proved for a continuous
method, that is, for the Cauchy problem for a differential equation, for
instance (\ref{cnm}), one can construct various finite difference schemes
for the solution of this Cauchy problem. These difference schemes give
discrete methods for the solution of equation (\ref{eq1}). For instance 
the methods of Euler and Runge-Kutta can be used. More detailed information 
about the applications and modifications of continuous Newton methods can 
be found in (\cite{g,zp,ap}).

The aim of this paper is to construct a continuous analog of iterative
scheme ~(\ref{eq3}), to prove a convergence theorem for this continuous analog
of (\ref{eq3}), and to test the method numerically by applying it to a
practically interesting nonlinear inverse problem of gravimetry.

The paper is organized as follows. In section 2 a continuous analog of
method (\ref{eq3}) is described and a convergence theorem for this method
is formulated. In section 3 this convergence theorem is proved. In
section 4 an inverse gravimetry problem is considered and the proposed
method is numerically tested. In our numerical experiments comparison of
different regularization functions
is done. Based on the results of the numerical experiments some recommendations
are given for the choice of the regularization function. 

\newpage

\bigskip
\noindent
{\bf 2.\ Continuous Gauss-Newton Method and Convergence Theorem}
\\$\left.\right.$
\setcounter{equation}{0}
\renewcommand{\theequation}{2.\arabic{equation}}
\vspace*{-0.5pt}
\noindent

In order to describe convergence rates we introduce the following

\medskip
\noindent
{\bf Definition 2.1.\ }
A positive function $\a(t)\in C^1 [0,\infty)$
is said to be a convergence rate function if $\a (t)$ decreases monotonically
to zero as $t\to \infty,$ $\a (t) \in C^1 [0,\infty)$ and $\ln \a (t)$ is concave,
that is, $\dot{\a}(t)/\a (t)$ is monotonically increasing.

\medskip
\noindent
{\bf Remark 2.2.\ }
The number $\a (0)$ can be chosen sufficiently large and simultaneously  
the number $|\dot{\a}(0)/\a(0)|$ can be sufficiently small.
Here and below the over dot denotes the derivative with respect to time
$\dot{x}:=dx/dt$.
For example, one can choose $\a(t)=b/(t+a)$, where $a$ and $b$ are positive
constants such that $a$ and $b/a$ are sufficiently large.

\vspace*{12pt}

A continuous analog of iterative process~(\ref{eq3}) is the following
Cauchy problem:
\be\label{eq4}
\dot{x}(t)=-[\ps(x(t))\pp(x(t))+\a(t)I]^{-1}[\ps(x(t))\p(x(t))+
\a(t)(x(t)-x_0)],
\ee
$$
x(0)=x_0.
$$
Denote by $\t{Ran}(L)$ the range of the linear operator $L$. The convergence
of the continuous analog of Gauss-Newton method (CAGNM) is established by
the following theorem, in which (and throughout this paper) the norms 
$||\pp(x)||$ and $||\ppp(x)||$ are the norms of linear and bilinear operators
from $H_1$ to $H_2$ and from $H_1\times H_1$ to $H_2$ respectively.  

\medskip
\noindent
{\bf Theorem 2.3.\ }
Let $\a(t)$ be a convergence rate function.
Assume that there exists a positive number $R$ for which Condition A
and the following conditions hold:
\begin{description}
\item[(i)] The Fr\'echet derivatives $\pp(x)$ and $\ppp(x)$ exist in the ball
$B(\hat{x},R)$ and satisfy the following inequalities:
\be\label{bound}
||\pp(x)|| \leq N_1, \q ||\ppp(x)|| \leq N_2 \q \forall x \in B(\hat{x},R),
\ee
where
$$ 
\f{\a(0)}{N_1N_2}\left(1-2N_1N_2||v||+\f{\dot{\a} (0)}{\a (0)}\right)\le R.
$$
\item[(ii)]
\be\label{iop}
x_0\in B(\hat{x},R)\cap[\hat{x}+\t{Ran}(\ps(\hat{x})\pp(\hat{x}))].
\ee
\item[(iii)] For some $v$, such that $\hat{x}-x_0=\ps(\hat{x})\pp(\hat{x})v$,
the following inequalities hold:
\be
1-2N_1N_2||v||+\f{\dot{\a} (0)}{\a (0)}>0, 
\ee
\be\label{in2}
\left(1-2N_1N_2||v||+\f{\dot{\a} (0)}{\a (0)}\right)^2-2N_1N_2||v||>0.
\ee
\end{description}
Then the following conclusions hold:
\begin{description}
\item[(i)] The solution $x=x(t)$ of problem (\ref{eq4}) exists, and
$x(t)\in B(\hat{x},R)$ for $t\in[0,\infty)$,
\item[(ii)] $||x(t)-\hat{x}||=O(\a (t))$ for $t \to \infty$.
\end{description}

\medskip
\noindent
{\bf Remark 2.4.\ }
Condition (ii) in Theorem 2.3 gives some restriction on the choice of 
an initial approximation point. It is not easy to verify this condition
algorithmically. However some kind of this condition is necessary if one works with 
equation (\ref{eq1}) with the operator $\ps(x)\pp(x)$, which is not boundedly 
invertible. If the operator $\ps\pp$ is injective but is not boundedly invertible,
then the image of the linear selfadjoint operator $\ps\pp$ is dense in $B(\hat{x},R)$
and consequently the set of the suitable initial approximation points satisfying
condition (ii) is also dense in $B(\hat{x},R)$. As our numerical results show 
(see section 4) the proposed method is practically efficient.

\bigskip
\noindent
{\bf 3.\ Proof of Theorem 2.3}
\\$\left. \right. $
\setcounter{equation}{0}
\renewcommand{\theequation}{3.\arabic{equation}}
\vspace*{-0.5pt}
\noindent

The main part of the proof is to show that the solution to problem (\ref{eq4}) does
not leave the ball $B(\hat{x},R)$ (Lemma 3.3).
In order to prove it, let us assume that
there exists such a point  $t_1 \in [0,\infty)$ that $x(t)$ intersects the
boundary of $B(\hat{x},R)$ for the first time at $t=t_1$. Hence  $x(t)$
belongs to the interior of the $B(\hat{x},R)$ for $t \in [0,t_1)$ and
$||x(t_1)-\hat{x}||=R$. Let us introduce an auxiliary function
\be\label{eq6}
w(t):=||x(t)-\hat{x}||/\a (t).
\ee
First, in Lemma 3.1, we derive a nonlinear differential inequality 
for $w(t)$. From
this differential inequality we get the estimate which shows that for all 
$t \in [0,t_1]$ the points of the integral curve of problem (\ref{eq4}) belong
to the interior of the ball $B(\hat{x},R)$. This contradiction proves that
the integral curve of the solution does not leave the above ball, and consequently
problem (\ref{eq4}) has
the global solution for $t\in [0,\infty)$. Also we show the boundedness
of the $w(t)$ and this implies, by formula (\ref{eq6}), strong convergence of
$x(t)$ to $\hat x$ for $t\to\infty$.

\vspace*{12pt}
\noindent
{\bf Lemma 3.1.\ } 
If the assumptions of Theorem 2.3 hold then the
differential inequality
\be\label{eq5}
\f{dw}{dt}\leq C_1w^2-C_2w+C_3,
\ee
is valid for $t \in [0,t_1]$, where
\be\label{eq11}
C_1=\f{N_1N_2}{2},\q  C_2=1-2N_1N_2||v||+\f{\dot{\a}(0)}{\a (0)},\q
C_3=||v||.
\ee

\vspace*{12pt}
\noindent
{\bf Proof.}
The G\^ateaux derivative $\p^{\pr\pr}(x,\xi_1,\xi_2)$ is a bilinear
operator such that
$$\pp(x+\xi_1)\xi_2-\pp(x)\xi_2:=\p^{\pr\pr}(x,\xi_1,\xi_2) +\eta\xi_2,
\hbox{ and }||\eta||\cdot||\xi_1||^{-1}\to 0 \hbox{ for }\xi_1\to 0,
||\xi_1||>0.$$
Let us define operators $K,G:B(\hat{x},R)\times H_1\times H_1\to H_2$
by the formulas:
\be\label{f1}
K(x,\xi_1,\xi_2)=\int\limits_0^1\int\limits_0^1\ppp(x+st\xi_1,\xi_1,\xi_2)
tdtds
\ee
and
\be\label{f2}
G(x,\xi_1,\xi_2)=\int\limits_0^1\ppp(x+t\xi_1,\xi_1,\xi_2)dt.
\ee
Then from (\ref{bound}) we get
\be\label{eq8}
|K(x,\xi_1,\xi_2)|\leq\f{N_2}{2}||\xi_1||\cdot||\xi_2||,\q
|G(x,\xi_1,\xi_2)|\leq N_2||\xi_1||\cdot||\xi_2||.
\ee
The following formulas will be used:
\be\label{eq7}
\p (\hat x)-\p (x)=\pp (x)(\hat x -x)+K(x,\hat x -x,\hat x -x),
\ee
and
\be\label{eq9}
(\pp (\hat x)- \pp(x))\xi=G(x,\hat x -x,\xi),
\ee
where $K$ and $G$ are defined by (\ref{f1}) and (\ref{f2})
respectively. Let us derive formulas (\ref{eq7}) and (\ref{eq9}).
One has
$$
\p(\hat x)-\p(x)=\il_0^1\f{d}{dt}\p(x+t(\hat x -x))dt
=\il_0^1\pp(x+t(\hat x -x))(\hat x -x)dt
$$
$$
=\pp(x)(\hat x -x)+\il_0^1[\pp(x+t(\hat x -x))-\pp(x)](\hat x -x)dt
$$
$$
=\pp(x)(\hat x -x)+\il_0^1 tdt\il_0^1 ds[\ppp(x+st(\hat x -x))](\hat x -x)
(\hat x -x)
$$
and
$$
[\pp(\hat x)-\pp(x)]\xi=\il_0^1\f{d}{dt}\pp(x+t(\hat x -x))\xi dt
=\il_0^1\ppp(x+t(\hat x -x),\hat x -x,\xi)dt.
$$

Since $\hat x$ solves (\ref{eq1}), one can rewrite equation
~(\ref{eq4}) as
$$
\f{dx}{dt}=-[\ps (x)\pp (x)+\a I]^{-1}[\ps (x)(\p (x)-\p(\hat
x))+\a (x-x_0)].
$$ 
From the condition (iii) of Theorem 2.3 and from (\ref{eq7})
we get
$$
\f{dx}{dt}=-[\ps (x)\pp (x)+\a I]^{-1}[-\ps(x)(\pp (x)(\hat x -x)+
K(x,\hat x -x,\hat x -x))+
$$ 
$$
\a (x-\hat x)+\a \ps (\hat x)\pp (\hat x)v],
$$
and therefore
$$ 
\f{dx}{dt}=   
-(x-\hat x )-[\ps (x)\pp (x)+\a I]^{-1}[-\ps (x)K(x,\hat x -x,\hat x -x)+
\a\ps (x)\pp (x)v+
$$
$$
\a (\ps(\hat x)\pp (\hat x)-\ps (x)\pp (x))v].
$$
Since $\hat x$ does not depend on t, it follows from (\ref{eq9}) that
$$
\f{d(x(t)-\hat x)}{dt}=\f{dx}{dt}=
-(x-\hat x )-[\ps (x)\pp (x)+\a I]^{-1}
[-\ps (x)K(x,\hat x -x,\hat x -x)
$$
$$
+\a\ps (x)\pp (x)v+\a(\ps(\hat x)-\ps (x))\pp(\hat x )v+
\a\ps(x)G(x,\hat x -x,v)].
$$
Now let us derive an inequality for $\f{d}{dt} ||x-\hat x ||^2.$ One has
$$
\f{d}{dt} ||x-\hat x ||^2=-2||x-\hat x ||^2+2([\ps (x)\pp (x)+\a I]^{-1}
[\ps(x)K(x,\hat x -x,\hat x -x)], x-\hat x)-
$$
$$
2\a([\ps (x)\pp (x)+\a I]^{-1}\ps (x)\pp(x)v,x-\hat x )+
2\a(\pp(\hat x)v,G(x,\hat x -x,[\ps (x)\pp (x)
$$
$$
+\a I]^{-1}(x-\hat x)))+
2\a([\ps (x)\pp (x)+\a I]^{-1}\ps(x)G(x,\hat x -x,v),x-\hat x  ).
$$
Since the operator $\ps (x)\pp (x)$ is selfadjoint and nonnegative
we have the following spectral representation:    
$$
[\ps (x)\pp (x)+\a I]^{-1}\ps (x)\pp(x)=\int_0^\infty\f{\l}{\l+\a}dE_\l,
$$
where $E_\l$ is the resolution of the identity of the selfadjoint  operator
$\ps(x)\pp(x)$.
Since $0\le\l/(\l+\a)\le 1$ for $\a>0$ and $\l\ge 0$, it follows that
\be\label{spes}
||[\ps (x)\pp (x)+\a(t)I]^{-1}\ps (x)\pp(x)||\le 1.
\ee
Also one has the following estimate:
\be\label{es1}
||[\ps (x)\pp (x)+\a(t)I]^{-1}||\le 1/\a(t).
\ee
From (\ref{spes}),(\ref{es1}) and (\ref{eq8}) 
one gets the following differential
inequality for $A(t):=||x(t)-\hat x||:$
$$
\dot{A}\leq -A+\f{N_1N_2}{2\a}A^2+\a ||v||+2N_1N_2 ||v|| A.
$$
In order to finish the proof of the Lemma 3.1, we
derive from the last inequality the inequality for $w(t)$ by taking into
account that $\dot{\a}(t)/\a (t)$ is monotonically increasing function.
$\Box $
  
The following lemma is a simple corollary of the more general results
established in (\cite{JS}).
\newpage

\vspace*{12pt}
\noindent
{\bf Lemma 3.2.\ }
Let $f(t,u)$ be a continuous function on
$[0,T]\times (-\infty,+\infty)$ such that the Cauchy problem
\be\label{eq12}
\dot u = f(t,u(t)),\q u(0)=u_0
\ee
is uniquely solvable on $[0,T]$ and $v(t)$ be a differentiable function
defined on $[0,T]$ and satisfies the conditions
\be\label{eq13}
\dot v \leq  f(t,v(t)),\q t\in[0,T],\q v(0)=v_0.
\ee
If $v_0\leq u_0$ then
$$v(t)\le u(t) \t{ for } t\in [0,T].$$
\vspace*{12pt}

It follows from inequality (\ref{in2}) that $c=\sqrt{C_2^2-4C_1C_3}>0$ for
constants $C_1,$ $C_2,$ $C_3$ defined in (\ref{eq11}).
Let $u_1$ and $u_2$ be correspondingly the smaller and the larger roots
of the equation $C_1u^2-C_2u+C_3=0$. For $u_0$ satisfying the
inequality $u_2<u_0<C_2/(2C_1)$ the solution of the Cauchy problem
\be\label{eq15}
\dot{u}=C_1u^2-C_2u+C_3,\q u(0)=u_0.
\ee
is given by the formula:
$$
\left\vert \f{u-u_2}{u-u_1}\right\vert
=\f{u_2-u_0}{u_0-u_1}e^{ct}.
$$
Let us show that  $u(t)$ is defined for all $t\in [0,\infty).$
Indeed, $u_0\in(u_1,u_2)$, so for sufficiently small $t$ one has
\be\label{eq17}
u=u_1+\f{u_2-u_1}{\f{u_2-u_0}{u_0-u_1}e^{ct}+1}.
\ee
Thus $u_1<u(t)<u(0)<u_2$.
This means that $u(t)$ does not leave the interval $(u_1,u_2)$ for all $t\in
[0,\infty)$
and $u(t)$ is well defined for all $t\in [0,\infty)$.

From condition (i) of Theorem 2.3 one obtains the following estimate:
$$
w(0)=\f{||x_0-\hat x ||}{\a (0)}<\f{1-2N_1 N_2 ||v||+\f{\dot{\a} (0)}{\a(0)}}{N_1
N_2}=\f{C_2}{2C_1}.
$$ 
Therefore from Lemmas 3.1 and 3.2 it follows that
\be\label{estim}
\f{||x(t_1)-\hat x ||}{\a (t_1)} \leq
u_1+\f{u_2-u_1}{\f{u_2-u_0}{u_0-u_1}e^{ct_1}+1}<u_0<\f{C_2}{2C_1}.
\ee

Thus 
$$
|| x(t_1)-\hat x ||<\f{C_2}{2C_1}\a(t_1)<\f{C_2}{2C_1}\a (0)\le R.
$$
This contradicts the assumption $|| x(t_1)-\hat x ||=R$. So the following lemma is
proved.
\newpage

\vspace*{12pt}
\noindent
{\bf Lemma 3.3.\ }
If the assumptions of Theorem 2.3 hold and for an arbitrary positive $T$ the solution
of the problem (\ref{eq4}) exists on the interval $[0,T]$, then the integral curve of
the solution of (\ref{eq4}) lies in the interior of the ball $B(\hat{x},R)$ for all
$t$ from the interval $[0,T]$.
\vspace*{12pt}

Now let us show that there exists the unique solution of (\ref{eq4}) on
$[0,\infty)$provided that $x_0$ satisfies conditions ii) and iii) of Theorem 2.3. The
Cauchy problem (\ref{eq4}) is equivalent
to the integral equation
\be\label{eq18}
x(t)=x_0+\il_0^t F(s,x(s))ds,
\ee
where
$$
F(s,x(s)):=-[\ps(x(s))\pp(x(s))+\a (s) I]^{-1}[\ps(x(s))\p(x(s))+\a (s) (x(s)-x_0)].
$$
Let us fix an arbitrary large positive number $T$ and use the successive
approximation method to solve equation (\ref{eq18}) on $[0,T]$:
\be\label{eq19}
x_{n+1}(t)=x_0+\il_0^t F(s,x_n (s))ds,\q x_0(t)=x_0,\q t\in [0,T].
\ee
Since $\pp(x)$ and $\ppp(x)$ are assumed to be bounded in $B(\hat{x},R)$, see
(\ref{bound}), and $\a(t)$ is positive on $[0,T]$, for every $t\in [0,T]$ the
function $F(t,x)$ has bounded Fr\'echet derivative with respect to $x$ in
$B(\hat{x},R)$. So one has:
$$||F(t,x_1)-F(t,x_2)||\le K(T)||x_1-x_2||$$
for all $t\in [0,T]$ and $x_1$, $x_2$ belong to $B(\hat{x},R)$.
Thus, one easily gets the estimate
$$
||x_{n+1}(s)-x_n (s) || \leq ||F(x_0)|| K^n(T)\f{T^n}{n!}
$$
valid on the maximal subinterval $[0,T_1]=\{t: t\in [0,T] \t{ and } x(t)\in
B(\hat{x},R)\}$. Therefore iterative process (\ref{eq19}) converges
uniformly and determines the unique solution of equation (\ref{eq18}) on $[0,T_1]$.
If $T_1<T$, it follows
from the maximality of the subinterval $[0,T_1]$ that $x(T_1)$ is a
boundary point
of $B(\hat{x},R)$. But this contradicts to Lemma 3.3. So the solution
of the problem (\ref{eq4}) exists and belongs to the interior of the ball
$B(\hat{x},R)$ on every interval $[0,T]$ and consequently on $[0,\infty)$.

To finish the proof of Theorem 2.3 it is sufficient to note that equation
(\ref{estim}) implies estimate
$$
||x(t)-\hat x || \leq \f{C_2}{2C_1}\a(t)
$$
for all $t\in[0,\infty)$.
$\Box$
\newpage

\bigskip
\noindent
{\bf 4.\ Numerical Results}
\\$\left. \right. $
\setcounter{equation}{0}
\renewcommand{\theequation}{4.\arabic{equation}}
\vspace*{-0.5pt}
\noindent

To test numerically the method described above, we chose the inverse gravimetry
problem (\cite{vas}). The goal of the numerical test is to illustrate the
choice of the regularization function $\a(t)$ and to compare two methods of
solving the Cauchy problem (\ref{eq4}): the Euler method, which corresponds to
the iterative scheme (\ref{eq3}), and the Runge-Kutta method.

Let the sources of a gravitational field with a constant density $\rho$ be
distributed in the domain
$$
D=\{-l\leq t \leq l,\q -H\leq z \leq -H+x(t)\},
$$
where $x(t)$ is an interface between two media, $l$ and $H$ are
parameters of the domain.
The potential $V$ of such a field is given by the double integral:
$$
V(t,z)=\frac{1}{2\pi}\int_D\int \rho\ln\frac{1}{\sqrt{(t-s)^2+
(z-\tau )^2}}dS =
$$
$$
=-\frac{\rho}{4\pi}\il_{-l}^l ds\il_{-H}^{-H+x(s)}\ln [(t-s)^2+
(z-\tau)^2]d\tau.
$$
For the $z$ - component of the gravitational field one has
$$
-\frac{\partial V(t,z)}{\partial  z} = -\frac{\rho}{4\pi}\il_{-l}^l
ds\il_{-H}^{-H+x(s)}
\frac{\partial   }{\partial   \tau}\ln [t-s)^2+(x-\tau)^2]d\tau =
$$
$$
=\frac{\rho}{4
\pi}\il_{-l}^l\ln\frac{(t-s)^2+(z+H)^2}{(t-s)^2+(z+H-x(s))^2} ds.
$$
In particular, on the surface $z=0$ we obtain the following nonlinear
operator equation
\be\label{eq26}
\p(x)\equiv\frac{\rho}{4\pi}\il_{-l}^lK(t,s,x(s))ds-y(t)=0,
\ee
where
$$
K(t,s,x(s))=\ln \frac{(t-s)^2+H^2}
{(t-s)^2+(H-x(s))^2}.
$$
The gravity strength anomaly $y(t)=-\frac{\partial  V(t,0)}{\partial z}$
is given and the interface between two media (with and without the sources
of a gravimetry field) $x(s)$ is to be determined.
  
Let $\p$ act between the pair of Hilbert spaces $H_1$ and $H_2.$ Assume that
$H_1=H^1 [-l,l]$ or $L_2 [-l,l]$ and $H_2 = L_2 [-l,l].$
The Fr\'echet derivative of this operator is the following one 
\be\label{eq27}
\pp(x)h=\il_{-l}^l \frac{2(H-x(s))h(s)}{(t-s)^2+(H-x(s))^2} ds.  
\ee
For any fixed $x\in \{x\in L_2 [-l,l],x\leq H-\varepsilon ,   
\varepsilon > 0\}$ the kernel
$$
K'_x(t,s,x(s))\equiv \frac{2(H-x(s))}{(t-s)^2+(H-x(s))^2}
$$
is a square integrable function on $[-l,l]\times[-l,l]$, therefore $\pp(x)$ 
in (\ref{eq27}) is a compact linear operator in $L_2[-l,l]$.
This means that the operators $\pp(x)$ and $\ps(x)\pp(x)$
are not boundedly invertible. So, one can not use classical iterative
schemes such as Newton or Gauss - Newton in the case of
equation (\ref{eq26}).

We solve the Cauchy problem (\ref{eq4}) for $\p$ given by (\ref{eq26})
with some regularization function $\a(t)$. The problem is numerically solved by means
of two finite difference methods, namely, Euler's method
\be\label{eq28}
x_{k+1}=x_k +\tau F(t_k,x_k),\q x_0=x(0)
\ee
and the Runge - Kutta method
\be\label{eq29}
x_{k+\frac{1}{2}} = x_k +\frac{\tau}{2} F(t_k,x_k),
\ee
$$   
x_{k+1} = x_k+\tau F(t_{k+\frac{1}{2}},x_{k+\frac{1}{2}}),\q x_0 = x(0).
$$
 Here
$$
F(t_k,x_k)\equiv -[\ps (x_k)\pp (x_k)+\a(t_k)I]^{-1}(\ps (x_k)\p (x_k)
+\a(t_k)(x_k-x_0))
$$
and an equal grid size $\tau > 0$ defines the node points,
$$
t_k=k\tau,\q k=0,1,..
$$
For the successful realization of Continuous Gauss-Newton method
an appropriate regularization function should be chosen. At the beginning
of the process values of $\a(t)$ should not be very small for the operator
$\ps(x(t))\pp(x(t))+\a(t)I$ to be stably invertible and at the
same time $\a(t)$ should tend to zero sufficiently fast to ensure
convergence of the function $x(t)$ to the solution of problem (\ref{eq26}).
In the numerical experiments the functions $\a_0/(\b+t)^m$, $\a_0e^{-\b t}$
and $\a_02^{-\b t}$ were used. The experiments have shown that
for all the considered functions the numerical solution is evaluated with
appropriate accuracy for sufficiently large range of the parameters
$m$, $\a_0$ and $\b$. The following tables illustrate the dependence of
the accuracy of the numerical results on parameters
for the following data $l=1,\q H=2, \q \rho =1,\q x_0=1$. For the numerical
tests the function $y(t)$ in (\ref{eq26}) was chosen as the solution
of the direct problem for the model function $x_{mod}(t)=(1-t^2)^2$. The integral in
(\ref{eq26}) was calculated by Simpson's formula with the number of node
points equals 201 and with a step size equal to $0.01$.

In the tables below $\Delta_E$ and $\Delta_R$ are the absolute errors,
$\sigma_E$ and $\sigma_R$ are the discrepancies $||\p(x(t))||$, see (\ref{eq26}),
of the Euler and the Runge-Kutta methods respectively. The first table shows the
dependence of the absolute errors and the discrepancies $\p(x)$ on the 
regularization function.

\newpage
   
\centerline{Table 1.}

\vspace{0.5cm}

\begin{tabular}{|c|c|c|c|c|c|}
\hline
\multicolumn{6}{|c|}{$\a_0=0.1,\q\tau=0.1$}\\
\hline
$\a(t)$ & N & $\Delta_E $ & $\sigma_E$ &$\Delta_R $ & $\sigma_R$\\
\hline
$\a_0(1+t)^{-2}$ & 38 & $0.26$  & $3.75\cdot10^{-2}$ & $0.27$
&$4.09\cdot10^{-2}$ \\
$\a_0(1+t)^{-4}$  & 25 & $0.25$  &$0.25$ & $0.11$
&$0.11$ \\
$\a_0(1+t)^{-6}$  & 43 &$0.12$   &$0.12$ &$1.75\cdot10^{-2}$ &
$2.03\cdot10^{-2}$ \\
$\a_0(1+t)^{-8}$ & 61 & $5.30\cdot10^{-2}$  & $2.87\cdot10^{-3}$
&$5.33\cdot10^{-2}$
&$3.58\cdot10^{-3}$ \\
$\a_0(1+t)^{-10}$ & 79 & $2.25\cdot10^{-2}$  & $4.90\cdot10^{-4}$
&$2.23\cdot10^{-2}$ & $6.22\cdot10^{-4}$ \\
$\a_02^{-3.5t}$ & 127 & $1.07\cdot10^{-2}$  & $1.14\cdot10^{-5}$
&$1.08\cdot10^{-2}$ & $3.03\cdot10^{-5}$ \\
$\a_0e^{-3.5t}$ & 85  & $1.08\cdot10^{-2}$ & $2.84\cdot10^{-4}$
& $1.08\cdot10^{-2}$ & $3.60\cdot10^{-4}$ \\
\hline
\end{tabular}

\vspace{1cm}

Then it is assumed that the type of a function $\a(t)$ is chosen and
the dependence of the absolute errors and the discrepancies on
parameters $\b$ (Tables 2 and 3) and $\a_0$ (Table 4) is analyzed.

\vspace{1.cm}

\centerline{Table 2.}

\vspace{0.5cm}

\begin{tabular}{|c|c|c|c|c|c|}
\hline
\multicolumn{6}{|c|}{$\a(t)=\a_0e^{-\b t},\q\a_0=0.1,\q\tau=0.1$}\\
\hline
$\b$ & N & $\Delta_E $ & $\sigma_E$ &$\Delta_R $ & $\sigma_R$\\
\hline
1  & 29  & 0.26  & $7.60\cdot10^{-2}$  &0.26  &$9.10\cdot10^{-2}$   \\
2  & 156 & $1.09\cdot10^{-2}$  &$4.58\cdot10^{-6}$  &$1.10\cdot10^{-2}$
&$1.52\cdot10^{-5}$
\\
3  &100  &$1.06\cdot10^{-2}$   &$5.78\cdot10^{-5}$  &$1.07\cdot10^{-2}$   &
$9.30\cdot10^{-5}$ \\
4  &73  &$1.15\cdot10^{-2}$   &$ 7.54\cdot10^{-4}$ &$1.14\cdot10^{-2}$
&$1.10\cdot10^{-3}$
\\
5  &56  &$1.52\cdot10^{-2}$   &$4.50\cdot10^{-3} $  &$1.50\cdot10^{-2}$   &
$6.12\cdot10^{-3}$ \\
6  &45  &$2.08\cdot10^{-2}$&$1.40\cdot10^{-2}$  &$2.17\cdot10^{-2}$
&$1.84\cdot10^{-2}$ \\
7  &37  &$3.90\cdot10^{-2}$  &$3.06\cdot10^{-2}$  &$3.89\cdot10^{-2}$  &
$3.80\cdot10^{-2}$  \\
8  &31  &0.11  &$5.82\cdot10^{-2}$  &$8.59\cdot10^{-2}$  & $6.52\cdot10^{-2}$ \\
9  &26  & 0.18 &$0.10$  &0.15  &0.11  \\
10 &23  &0.24  &0.14  &0.20  & 0.16 \\
\hline
\end{tabular}

\newpage

\centerline{Table 3.}

\vspace{0.5cm}

\begin{tabular}{|c|c|c|c|c|c|}
\hline  
\multicolumn{6}{|c|}{$\a(t)=\a_0e^{-\b t},\q\a_0=0.1,\q\tau=0.6$}\\
\hline
$\b$ & N & $\Delta_E $ & $\sigma_E$ &$\Delta_R $ & $\sigma_R$\\
\hline
1  & 4  & 0.23  & $5.35\cdot10^{-2}$  &0.26  &0.12   \\
2  & 20& $4.13\cdot10^{-2}$  &$1.00\cdot10^{-3}$  &$1.09\cdot10^{-2}$
&$6.11\cdot10^{-5}$
\\
3  &16  &$8.90\cdot10^{-2}$   &$5.04\cdot10^{-2}$  &$1.05\cdot10^{-2}$   &
$1.73\cdot10^{-4}$ \\
4  &12  &$1.46\cdot10^{-2}$   &$ 1.29\cdot10^{-4}$ &$1.25\cdot10^{-2}$
&$1.79\cdot10^{-3}$
\\
5  &9  &$1.76\cdot10^{-2}$   &$7.14\cdot10^{-4} $  &$1.84\cdot10^{-2}$   &
$1.08\cdot10^{-2}$ \\
6  &7  &0.16&$3.85\cdot10^{-3}$  &$2.98\cdot10^{-2}$
&$3.22\cdot10^{-2}$ \\
7  &6 &$0.38$  &$2.58\cdot10^{-2}$  &$0.12$  &
$5.61\cdot10^{-2}$  \\
8  &5  &0.83  &$1.48$  &$0.18$  & $8.60\cdot10^{-2}$ \\
9  &4  & 0.83 &$01.48$  &0.21  &0.15  \\
10 &3  &0.83  &01.48  &0.23  & 0.19 \\
\hline
\end{tabular}

\vspace{1.cm}

\centerline{Table 4.}

\vspace{0.5cm}

\begin{tabular}{|c|c|c|c|c|c|}
\hline
\multicolumn{6}{|c|}{$\a(t)=\a_0e^{-\b t},\q\b=3.5,\q\tau=0.1$}\\
\hline
$\a_0$ & N & $\Delta_E $ & $\sigma_E$ &$\Delta_R$ & $\sigma_R$\\
\hline
$10^{-3}$  & 71  & $1.62\cdot10^{-2}$  & $9.77\cdot10^{-4}$  &$1.61\cdot10^{-2}$
&$1.33\cdot10^{-2}$ \\
$10^{-2}$  & 79& $1.33\cdot10^{-2}$  &$4.55\cdot10^{-4}$
&$2.29\cdot10^{-2}$ &$2.90\cdot10^{-4}$
\\
$10^{-1}$  &85  &$1.07\cdot10^{-2}$   &$2.84\cdot10^{-4}$
&$1.08\cdot10^{-2}$ &
$3.60\cdot10^{-4}$ \\
\hline
\end{tabular}

\vspace{1cm}

Analyzing the results of the numerical experiments (a part of them is included
in the Tables) one concludes the following:
\begin{description}
\item[(i)] The Runge-Kutta method is more stable with respect to changes of
the regularization function $\a(t)$ and the step size $\tau$, than the Euler method;
\item[(ii)] for all the considered functions suitable parameters can be chosen,
however in the case when $\a(t)=\a_0e^{-\b t}$ the accuracy with which the solution
$x(t)$ is calculated is higher;
\item[(iii)] an appropriate range of values of the parameter $\a_0$ is from
0.001 to 0.1, for larger values the accuracy is lower, and for smaller values
the processes do not converge;
\item[(iv)] the range of appropriate values of $\b$ is large enough: from 2 to
6 for $\tau=0.6$ and from 2 to 7 for $\tau=0.1$.
\end{description}
\newpage

\medskip
\noindent
{\bf Remark 4.1.\ }
The reason why result (i), formulated above, is emphasized can
be understood if one remembers that problem (\ref{eq26}) is ill-posed.
If a problem is ill-posed then the usage of a higher-order accuracy difference scheme
(or quadrature formula) may lead to less accurate results, as was observed in the
literature (see e.g. (\cite{mg}), p. 155). On the other hand, if a problem is
well-posed, then the usage of a higher-order accuracy scheme should lead to more
accurate results.

\bigskip
\noindent
{\bf Acknowledgments}
\\$\left. \right. $
\setcounter{equation}{0}
\renewcommand{\theequation}{4.\arabic{equation}}
\vspace*{-0.5pt}

The authors thank Dr. V.Protopopescu for useful remarks.

\newpage

\nonumsection{References}

\end{document}